\documentclass[]{llncs}  

\newcommand{\verticalText}[1]{\rotatebox[origin=c]{90}{#1}}
 
\usepackage{amsmath,amsfonts,amssymb}
\usepackage{graphicx}
\usepackage[colorlinks=true, allcolors=blue]{hyperref}
\usepackage{subcaption}
\expandafter\def\csname ver@subfig.sty\endcsname{}

\usepackage{hyperref}
\usepackage{amsmath} 
\usepackage{siunitx}
\usepackage{svg}
\usepackage{subfig}
\usepackage{pgf}
\usepackage{xfrac}
\usepackage{xcolor}
\usepackage{mathtools}
\usepackage{bm}
\usepackage{multirow}
\usepackage{longtable}
\usepackage{tablefootnote}
\usepackage{graphbox}
\usepackage{adjustbox}
\usepackage{listings}
\usepackage{booktabs}
\usepackage{amssymb}
\usepackage[T1]{fontenc}
\usepackage{graphicx}

\definecolor{tum_blue_dark_2}{HTML}{0e396e}
\definecolor{tum_blue_light}{HTML}{5e94d4}
\definecolor{tum_blue_light_2}{HTML}{c2d7ef}
\definecolor{tum_red}{HTML}{ea7237}
\definecolor{tum_orange}{HTML}{f7b11e}
\definecolor{tum_yellow}{HTML}{fed702}
\definecolor{tum_grey_5}{HTML}{abb5be}
\definecolor{tum_grey_1}{HTML}{20252a}
\definecolor{tum_green}{HTML}{9fba36}

\title{Redefining spectral unmixing for in-vivo brain tissue analysis from hyperspectral imaging}

\begin{document} 

\author{
Martin Hartenberger \inst{1} \and
Huzeyfe Ayaz \inst{1} \and
Fatih Ozlugedik \inst{1} \and
Charly Caredda \inst{2,4} \and
Luca Giannoni \inst{3} \and
Frédéric Lange \inst{3} \and
Laurin Lux \inst{1} \and
Jonas Weidner \inst{1} \and
Alex Berger \inst{1} \and
Florian Kofler \inst{1} \and
Martin Menten \inst{1} \and
Bruno Montcel \inst{2,4} \and
Ilias Tachtsidis \inst{3} \and
Daniel Rueckert \inst{1,5,6} \and
Ivan Ezhov \inst{1}
}

\institute{
Technical University of Munich (TUM) and TUM University Hospital \and
Institut National des Sciences Appliquées de Lyon, Univ. Claude Bernard Lyon 1, Univ. de Lyon \and
University College London \and Univ. Jean Monnet Saint-Etienne, Institut National de la Santé et de la Recherche Médicale, CNRS \and Department of Computing, Imperial College London \and Munich Center for Machine Learning (MCML)
   }

\maketitle

\begin{abstract}



In this paper, we propose a methodology for extracting molecular tumor biomarkers from hyperspectral imaging (HSI), an emerging technology for intraoperative tissue assessment. To achieve this, we employ spectral unmixing, allowing to decompose the spectral signals recorded by the HSI camera into their constituent molecular components. Traditional unmixing approaches are based on physical models that establish a relationship between tissue molecules and the recorded spectra. However, these methods commonly assume a linear relationship between the spectra and molecular content, which does not capture the whole complexity of light-matter interaction. To address this limitation, we introduce a novel unmixing procedure that allows to take into account non-linear optical effects while preserving the computational benefits of linear spectral unmixing. 
We validate our methodology on an in-vivo brain tissue HSI dataset and demonstrate that the extracted molecular information leads to superior classification performance.
\end{abstract}


\section{Introduction}
Diffuse gliomas, constituting approximately 80$\%$ of malignant primary brain tumors, remain challenging to treat due to their aggressive growth and the lack of a clear border between the tumor and healthy tissue. 
The standard treatment involves surgical resection, but current imaging techniques often lack the precision for optimal surgical navigation.

Hyperspectral imaging (HSI) has emerged as a potential alternative for living tissue assessment \cite{lin2016probe,studier2023heiporspectral,li2023spatial,ezhov2024learnable}. HSI systems operate using non-ionizing light from the visible to near-infrared (NIR) range, measuring reflected light from the same surface across hundreds of different wavelengths. The light-matter interaction can be considered to be governed by two fundamental physical processes: molecular (or, more precisely, chromophore) absorption and scattering. 



Different computational methods exist to connect the intensity of the reflected light with the molecular composition of living tissue, i.e. solve the spectral unmixing problem. The effect of incoming light on the tissue can be mathematically described using the Beer-Lambert law (BLL):

\begin{equation}
I_R(\lambda) = I_0(\lambda) e^{-(\mu_a(\lambda) + \mu_s(\lambda)) \cdot l}    
\end{equation}

Here, $I_0(\lambda)$ represents the intensity of the incident light, while $I_R(\lambda)$ is the intensity of the reflected light detected by the camera. The terms $\mu_a(\lambda)$ and $\mu_s(\lambda)$ correspond to the absorption and scattering coefficients, respectively, $\lambda$ denotes the wavelength, and $l$ is the optical pathlength.
The absorption coefficient $\mu_a(\lambda)$ is, in the simplest case, defined as a linear sum over absorption characteristic of the considered tissue molecules $\mu_a(\lambda) = \sum_i c_i m_a^i(\lambda)$, weighted by a molecular concentration $c_i$. 
The scattering coefficient, $\mu_s(\lambda)$, can be expressed as $\mu_s \sim a\lambda^{-b}$, where $b$ denotes the degree of the power-law model and $a$ is the scaling coefficient \cite{jacques_optical_2013}. By matching the measured reflected intensity with the one modelled by the Beer-Lambert law, we can estimate the optimal values for the molecular concentration.  

One of the main challenges for solving the spectral unmixing problem is identifying the set of endmembers (e.g., molecules) shaping the reflection spectrum. If one selects a set of molecules different from the tissue-relevant one, the wrongly inferred molecular concentrations, as a result, might not be useful for tissue analysis tasks. Another strong assumption that is made in Eq. 1 is the (logarithmically) linear relation between the intensity and the physical phenomena behind the dissipation of incoming light energy, i.e. absorption and scattering:

\begin{equation}
I_A = -\frac{1}{l}ln\frac{I_R(\lambda)}{I_0(\lambda)} = \textbf{M}\boldsymbol{\alpha}     
\end{equation}

Here, $\textbf{M}$ is a matrix of absorption and scattering spectra, and $\boldsymbol{\alpha}$ is a vector of weights (molecular concentrations $c$ and scattering weight $a$). Even though it is substantially simpler to solve the linear system compared to non-linear counterparts\footnote{Linear systems often have deterministic, closed-form solutions, can be convex and be solved using e.g. the pseudoinverse.}, the physical reality of the light-matter interaction can be far from linear (multiple scattering events, re-absorption of scattered photons, wavelength- and tissue-dependent pathlength, etc., can cause the measured intensity to deviate from the assumed linear behavior). Thus, the use of linear descriptions can potentially distance us further from accurate molecular predictions.

Our contributions are:

1. We design a procedure to implicitly include the nonlinear character of light-matter interaction into the linear unmixing. We achieve this by redefining the unmixing problem via introduction of pseudo-endmembers extracted from Monte-Carlo simulations of light-matter interaction.

2. We show how the proposed procedure can improve targeted endmember detection methods, such as Orthogonal Subspace Projection (OSP) and Constrained Energy Minimization (CEM). During unmixing, these methods allow to extract contributions to the reflection spectrum from a selected subset of molecules of interest but can be suboptimal in extracting plausible molecular maps.   

3. We qualitatively and quantitatively analyze the utility of the method on the brain tissue HSI dataset and demonstrate its practical benefit for the downstream brain tissue analysis task.

\section{Methods}

First, we introduce the standard targeted detection approaches. Then, we demonstrate how these approaches can be enhanced by means of the pseudo-endmember spectra derived from Monte-Carlo simulations.

\noindent\textbf{2.1. Orthogonal subspace projection.}

Using OSP for endmember detection was first proposed by Harsanyi et al. \cite{harsanyi_hyperspectral_1994}. The main idea behind the technique is that assuming the spectrum $I_A$ is made of a linear mixture of $n$ endmembers $m_i$, the contribution of endmembers can be entirely removed by projecting the reflection spectrum onto their orthogonal subspace.
Dividing the endmembers into a target endmember $\textbf{t}=m_{n}$ and the remaining endmembers ${m_1,m_2,\dots,m_{n-1}}$, the contribution of $\textbf{t}$ can be extracted using the projection matrix
\begin{equation}
    P^\perp_{U}=\mathbb{Id}- \textbf{U} \textbf{U}^\mathsf{*}
    \label{equ:proj_multi}
\end{equation}
\begin{equation}
    P^\perp_{U}I_A=P^\perp_{U} \textbf{M}\boldsymbol{\alpha} =(\mathbb{Id}- \textbf{U} \textbf{U}^\mathsf{*})\textbf{M}\boldsymbol{\alpha}
    \label{equ:proj_multi}    
\end{equation}
\begin{equation}
    P^\perp_{U}I_A = (\mathbb{Id}- \textbf{U} \textbf{U}^\mathsf{*})\textbf{t}\alpha_t + (\mathbb{Id}- \textbf{U} \textbf{U}^\mathsf{*})\textbf{U}\boldsymbol{\alpha_U}
    \label{equ:proj_multi}    
\end{equation}
\begin{equation}
    P^\perp_{U}I_A = (\mathbb{Id}- \textbf{U} \textbf{U}^\mathsf{*})\textbf{t}\alpha_t 
    \label{equ:proj_multi}    
\end{equation}

where the columns of $\textbf{U}$ form a span  $[m_1, m_2, \dots, m_{n-1}]$, and $\textbf{U}^*$ is a pseudoinverse.  
The last term in Eq. 5 is cancelled out since by definition of a pseudoinverse $\textbf{U} \textbf{U}^\mathsf{*}\textbf{U}=\textbf{U}$. 
To improve the signal-to-noise ratio (SNR), the inner product between a vector $\textbf{t}$ and the projected spectrum can be taken as a second step (it can be shown \cite{harsanyi_hyperspectral_1994} that the SNR of $x^\mathsf{T}P^\perp_{U} I_A$ is maximized for $x = \textbf{t}$).

Applying the OSP operator $\textbf{t}^\mathsf{T} P^\perp_{U}$ to each pixel of a preprocessed HSI, ${I}_A \in \mathbb{R}^{X\times Y\times k}$ ($X$ and $Y$ refer to spatial dimensions while $k$ to the spectral dimension), one obtains a heatmap $H^{OSP}_t \in \mathbb{R}^{X \times Y}\ $ for the target endmember $\textbf{t}$:
\begin{equation}
    H^{OSP}_t(x,y) = \textbf{t}^\mathsf{T} P^\perp_{U} {I}_A(x,y)\quad \forall x\in[X], y\in[Y].
    \label{equ:osp_heatmap}
\end{equation}

\noindent\textbf{2.2. Constrained energy minimization.}

Like OSP, CEM is a technique for extracting information about a specific target endmember $\textbf{t}$ \cite{hsuan_ren_comparison_2003}.
However, unlike for OSP, only the target endmember spectrum itself is needed to obtain its heatmap.
This has the advantage that no assumptions about the other endmembers present in the mixture have to be made, and the method typically demonstrates superior performance in effectively removing unidentified signal sources and reducing noise \cite{qian_du_comparative_2003}. 

CEM aims to find a filter $w_t\in \mathbb{R}^k$, which minimizes the energy contribution of all other endmembers while preserving the energy contribution from the target endmember $\textbf{t}$:

\begin{equation}
    E(w) = \sum_{x=0}^X \sum_{y=0}^Y (w^{} I_A)^2
\end{equation}

\[
\begin{cases}
\min_{w_t} E(w) \\
\text{subj. to } \textbf{t}^T w = 1.
\end{cases}
\]

Assuming the same linear model as for OSP, the optimal filter for this optimisation objective can be shown to be \cite{hsuan_ren_comparison_2003}:
\begin{equation}
    w^{CEM}_t = \frac{R^{-1}\textbf{t}}{\textbf{t}^\mathsf{T}R^{-1}\textbf{t}},
    \label{equ:cem_solution}
\end{equation}
where R is the auto-correlation matrix of $I_A$:
\begin{equation}
    R = \frac{1}{X*Y}\sum_{x=0}^X \sum_{y=0}^Y {I}_A(x,y) {I}_A(x,y)^\mathsf{T}.
    \label{equ:R_matrix}
\end{equation}

Note that the optimization is performed over all pixels, i.e. the filter is applied to the whole HSI image. Since the standard CEM is susceptible to noise, Wu et al. \cite{wu_constrained_2022} have proposed improved constrained energy minimization (ICEM), where a regularization term is added to the auto-correlation matrix. 
The equation for the filter vector in ICEM is
\begin{equation}
    w^{ICEM}_t = \frac{(R+\kappa\mathbb{Id})^{-1}\textbf{t}}{\textbf{t}^\mathsf{T}(R+\kappa\mathbb{Id})^{-1}\textbf{t}}.
    \label{equ:icem_solution}
\end{equation}
where $\kappa$ is the regularization strength.

Based on the filtering vectors obtained by Eq. \ref{equ:icem_solution}, molecular heatmaps for ${I}_A$ and target endmember $\textbf{t}$ can be calculated using the following formula:
\begin{equation}
    H^{ICEM}_t(x,y) = (w^{ICEM}_t)^\mathsf{T} I_A(x,y) = \frac{\textbf{t}^\mathsf{T}(R+\kappa\mathbb{Id})^{-1}I_A(x,y)}{\textbf{t}^\mathsf{T}(R+\kappa\mathbb{Id})^{-1}\textbf{t}} \quad \forall x\in[X], y\in[Y].
\end{equation}

\noindent\textbf{2.3. Monte Carlo derived pseudo-endmembers.}

For techniques such as OSP and ICEM to produce good results using the endmember spectra from the literature, the linear mixture assumption must hold. However, the physical nature of light-matter interactions can be highly non-linear. As a result, relying on linear descriptions may lead to inaccurate molecular inference.
A solution we propose here is to extract the impact of non-linear effects caused by the change of molecular concentration from a more realistic reflectance model, e.g. the one based on a Monte Carlo simulation of light-matter interaction, which takes non-linear effects into account. 
Based on this model, a linearization around the typical chromophore concentration in brain tissue can be performed, resulting in a new linear model.

Let $I_A = I_A(\mathbf{c},a)$ be the light attenuation obtained using the nonlinear model, where 
$\mathbf{c}= [c_1,c_2,\dots,c_n]$ is the vector of molar concentrations, and $a$ is the scattering weight of the assumed scattering model.
The partial derivative of $I_A$ with respect to the concentration $c_1$ can be approximated by
\begin{equation}
    \frac{\partial I_A}{\partial c_1} \approx \frac{1}{\Delta c_1} \left[ I_A\left(c_1 + \Delta c_1,c_2,\dots,c_n ,a \right) - I_A\left(c_1,c_2,\dots,c_n,a \right)\right] = m_1'.
    \label{equ:r_diff_partial_c}
\end{equation}
The gradient of $I_A$ with respect to the whole concentration vector $\mathbf{c}$ (and similarly for the scattering weight $a$) can therefore be approximated by
\begin{equation}
    \frac{ \partial I_A}{\partial \textbf{c}}^\mathsf{T} \approx [m_1', m_2', \dots, m_n'].
\end{equation}

and thus, in the first approximation, the partial derivatives can be interpreted as endmembers for the following linear model: 
\begin{equation}
    \Delta I_A = M'\Delta\mathbf{p} 
    \label{equ:lmm_mc}
\end{equation}
where $\Delta I_A = I_A - I_A^{ref}$ is the relative spectrum with respect to a reference spectrum $I_A^{ref}$ with fixed endmember abundances,  $M'=[m_1', m_2', \dots, m_n',m_{scatter}']$, and $\Delta\mathbf{p}=[\Delta c_1, \Delta c_2,\dots, \Delta c_n,\Delta a]$ is the unknown differential abundance vector. 

In simple terms, our approach implies, first, computing the change of light attenuation upon the change of abundance of a particular endmember according to a realistic non-linear model. After computing it individually for every endmember, we then use these changes of attenuation $m'_i=\frac{\partial I_A}{\partial p_i}$ to define a new set of pseudo-endmembers $M'$. 


For the nonlinear model, we selected a Monte Carlo simulator that stochastically simulates light propagation in matter from \cite{wang1995mcml,alerstam2008parallel}.
The simulations are performed using a single-layer model with near-infinite thickness and width. An anisotropy factor of $g=0.9$ and absolute refractive index $n=1.4$ are assumed following the literature \cite{jacques_optical_2013}. 
For the scattering coefficient, the power-law model for Rayleigh scattering with $b=1.3$ is assumed \cite{jacques_optical_2013}. 

\noindent\textbf{2.4. Dataset and evaluation setup.}

For our experiments, we used a publicly available HSI dataset of in-vivo brain tissue \cite{university_of_las_palmas_de_gran_canaria_hsi_nodate}. The spectral range of the used HSI system goes from 400 to 1000 \si{\nano\meter}.
Across the spectral range, 826 bands are recorded with a uniform step size and bandwidth of 2 to 3 \si{\nano\meter}. The dataset also includes labels for four classes of pixels: Normal tissue (\textit{NT}), tumor tissue (\textit{TT}), blood vessels (\textit{BV}), and background (\textit{BG}). Over the whole dataset, images were very sparsely labeled for the \textit{NT} (300339 pixels), \textit{TT} (21251), \textit{BV} (98783) and \textit{BG} class (205467), see Fig. \ref{fig:classification_maps} for an example of a labeled HSI image.

For endmembers, we considered typical absorbing chromophores of brain tissue, including water, lipids, deoxyhemoglobin (HHb), oxyhemoglobin ($\text{HbO}_2$), and cytochromes (Cytc, Cytb, and cytochrome-c-oxidase (CCO)), as well as scattering. The three cytochromes each exist in an oxidized (ox) and reduced state (red). Absorption properties of these molecules can be found in \cite{yu_functional_2020}. The scattering spectrum is estimated analogously to the MC simulations with the Rayleigh power-law model.

We perform a quantitative evaluation by comparing the tissue classification performance of different models using preprocessed HSI data only and utilizing both HSI and heatmap data ($H^{OSP}_t(x,y), \ H^{ICEM}_t(x,y)$) including the proposed ones derived from the MC simulations.
The premise is that heatmaps generated through OSP and ICEM are not merely extracted from the HSI images but obtained by considering the absorption and scattering properties of the tissue. Thus, we hypothesize that such maps introduce a form of inductive bias that should enhance the model's classification performance. The MLP architecture is chosen for all three of the classification models since MLP is a common choice \cite{fabelo2019deep,martin2024machine} for analysing the HSI data from the Helicoid dataset.

\section{Results and Discussion}

\begin{figure}[t]
    \centering
        \subfloat[]
            [\centering Using endmembers $M$ from the literature.]
            {
            \includegraphics[width=0.9\linewidth]{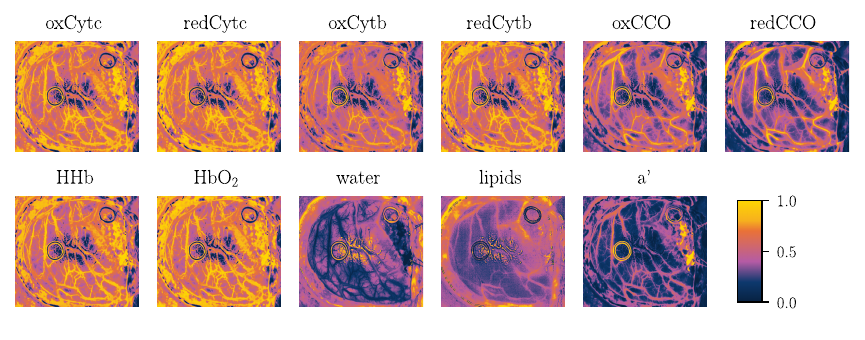}
            }
        \\
        \subfloat[]
            [\centering Using pseudo-endmembers $M'$ from MC simulations.]
            {
            \includegraphics[width=0.9\linewidth]{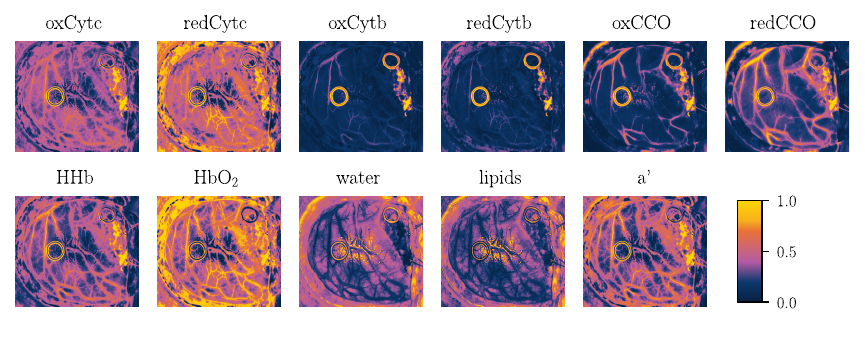}        
            }
    \caption{Differential heatmaps for HSI (12-1) obtained using ICEM ($\kappa=1$).}
        \label{fig:icem_rel_heatmaps_012}
\end{figure}

For all methods, we analysed absolute and differential molecular heatmaps. 
For the differential heatmaps, before performing OSP or ICEM, a spectrum belonging to a randomly selected pixel is subtracted from all other spectra of the image. For the absolute heatmaps, we analysed the spectra $I_A$ directly. In Fig. \ref{fig:icem_rel_heatmaps_012}, one can see an example of ICEM maps inferred from a typical HSI image. We see that the ICEM heatmaps (with $\kappa=1$ that appeared least noisy) for the cytochromes, hemoglobin, and scattering exhibit a high similarity. All of their heatmaps equally highlight vascular structures and the tumor region, while according to our prior biological knowledge, the molecules should highlight semantically different tissue areas. A different pattern is observed though when we base the spectral unmixing on the pseudo-endmembers obtained from the MC simulations. Cytochromes, the metabolic molecules, clearly outline the proliferating tumor area, while hemoglobins, the main molecules of blood, contrast the vascular tree.

In addition, we quantitevly evaluate our approach on the tissue classification task. 
For the first tissue classification model, the input vector is the pixel HSI spectrum - baseline BL model. For the second model, an input vector for each pixel is constructed by concatenating all the heatmap values obtained for that pixel from both the OSP and ICEM techniques - heatmap-only OHM model. 
Another HM model combines the feature vector from the BL and OHM models into one large feature vector.
Finally, NMC represents a model that inputs a vector analogous to the HM model but without the MC-derived heatmaps.



The provided summary of the class-wise metrics in Tab. 1 shows the performance differences among all models — BL, HM, OHM, and NMC — across various evaluation metrics. The HM model exhibits the most consistent performance across different metrics (with up to a 7\% increase compared to the baseline BL method), particularly with higher F1 scores across all semantic classes. The NMC model generally falls between the BL and HM models in terms of overall performance, indicating the importance of the proposed MC-derived heatmaps.
\begin{table}[t]
 \caption{Mean and standard deviation of performance metrics (accuracy, precision, recall, specificity and F1 score) over test images across five folds. 
 }
    \label{tab:metrics_2x2}
    \centering
    \begin{tabular}{lcccc|lcccc}
        \toprule
        \multicolumn{5}{c|}{\textbf{Normal tissue (NT)}} & \multicolumn{5}{c}{\textbf{Tumor tissue (TT)}} \\
        \midrule
        & BL & HM & OHM & NMC &  & BL & HM & OHM & NMC \\
        \midrule
        Acc.    & 0.89$_{\pm0.04}$  & \textbf{0.92}$_{\pm0.04}$  & 0.90$_{\pm0.02}$  & 0.91$_{\pm0.05}$  &     & 0.92$_{\pm0.02}$  & \textbf{0.93}$_{\pm0.02}$  & 0.92$_{\pm0.03}$  & 0.91$_{\pm0.03}$  \\
        Prec.   & 0.82$_{\pm0.09}$  & \textbf{0.89}$_{\pm0.01}$  & 0.85$_{\pm0.08}$  & 0.87$_{\pm0.11}$  &    & 0.51$_{\pm0.22}$  & \textbf{0.52}$_{\pm0.26}$  & 0.38$_{\pm0.26}$  & 0.41$_{\pm0.27}$  \\
        Rec.      & \textbf{0.94}$_{\pm0.03}$  & 0.93$_{\pm0.03}$  & 0.91$_{\pm0.03}$  & 0.90$_{\pm0.1}$  &       & 0.32$_{\pm0.17}$  & \textbf{0.35}$_{\pm0.21}$  & 0.17$_{\pm0.12}$  & 0.32$_{\pm0.24}$  \\
        Spec. & 0.87$_{\pm0.06}$  & \textbf{0.93}$_{\pm0.06}$  & 0.90$_{\pm0.05}$  & 0.92$_{\pm0.06}$  &  & 0.98$_{\pm0.01}$  & 0.98$_{\pm0.01}$  & \textbf{0.99}$_{\pm0.02}$  & 0.97$_{\pm0.04}$  \\
        F1    & 0.85$_{\pm0.06}$  & \textbf{0.90}$_{\pm0.05}$  & 0.86$_{\pm0.03}$  & 0.87$_{\pm0.09}$  &     & 0.34$_{\pm0.13}$  & \textbf{0.37}$_{\pm0.20}$  & 0.22$_{\pm0.12}$  & 0.33$_{\pm0.23}$  \\
        \midrule
        \multicolumn{5}{c|}{\textbf{Blood vessels (BV)}} & \multicolumn{5}{c}{\textbf{Background (BG)}} \\
        \midrule
        &BL & HM & OHM & NMC &  & BL & HM & OHM & NMC \\ 
        \midrule
        Acc.    & \textbf{0.96}$_{\pm0.01}$  & 0.95$_{\pm0.01}$  & 0.95$_{\pm0.01}$  & 0.96$_{\pm0.02}$  &     & 0.91$_{\pm0.03}$  & \textbf{0.93}$_{\pm0.04}$  & 0.93$_{\pm0.03}$  & \textbf{0.93}$_{\pm0.03}$  \\
        Prec.   & 0.80$_{\pm0.10}$  & 0.80$_{\pm0.18}$  & \textbf{0.86}$_{\pm0.18}$  & 0.82$_{\pm0.16}$  &    & \textbf{0.97}$_{\pm0.03}$  & 0.96$_{\pm0.04}$  & 0.94$_{\pm0.07}$  & 0.95$_{\pm0.05}$  \\
        Rec.      & 0.83$_{\pm0.09}$  & \textbf{0.87}$_{\pm0.05}$  & 0.83$_{\pm0.10}$  & 0.85$_{\pm0.07}$  &       & 0.82$_{\pm0.07}$  & \textbf{0.88}$_{\pm0.09}$  & 0.87$_{\pm0.08}$  & 0.86$_{\pm0.08}$  \\
        Spec. & \textbf{0.98}$_{\pm0.01}$  & 0.96$_{\pm0.01}$  & 0.97$_{\pm0.02}$  & 0.97$_{\pm0.01}$  &  & \textbf{0.97}$_{\pm0.03}$  & 0.96$_{\pm0.03}$  & 0.95$_{\pm0.02}$  & 0.95$_{\pm0.02}$  \\
        F1    & 0.79$_{\pm0.08}$  & \textbf{0.80}$_{\pm0.12}$  & 0.80$_{\pm0.09}$  & 0.80$_{\pm0.10}$  &    & 0.86$_{\pm0.04}$  & \textbf{0.89}$_{\pm0.04}$  & 0.88$_{\pm0.06}$  & 0.88$_{\pm0.04}$  \\
        \bottomrule
    \end{tabular}
\end{table}

\begin{figure}[t]
    \centering
    \subfloat{
        \centering
        \begin{adjustbox}{valign=t}
            \setlength{\tabcolsep}{1pt}
            \begin{tabular}{@{} *{4}{c} @{}}
                & sRGB & BL Model & HM Model\\
                \verticalText{15-1} & \includegraphics[align=c, width=1.8cm]{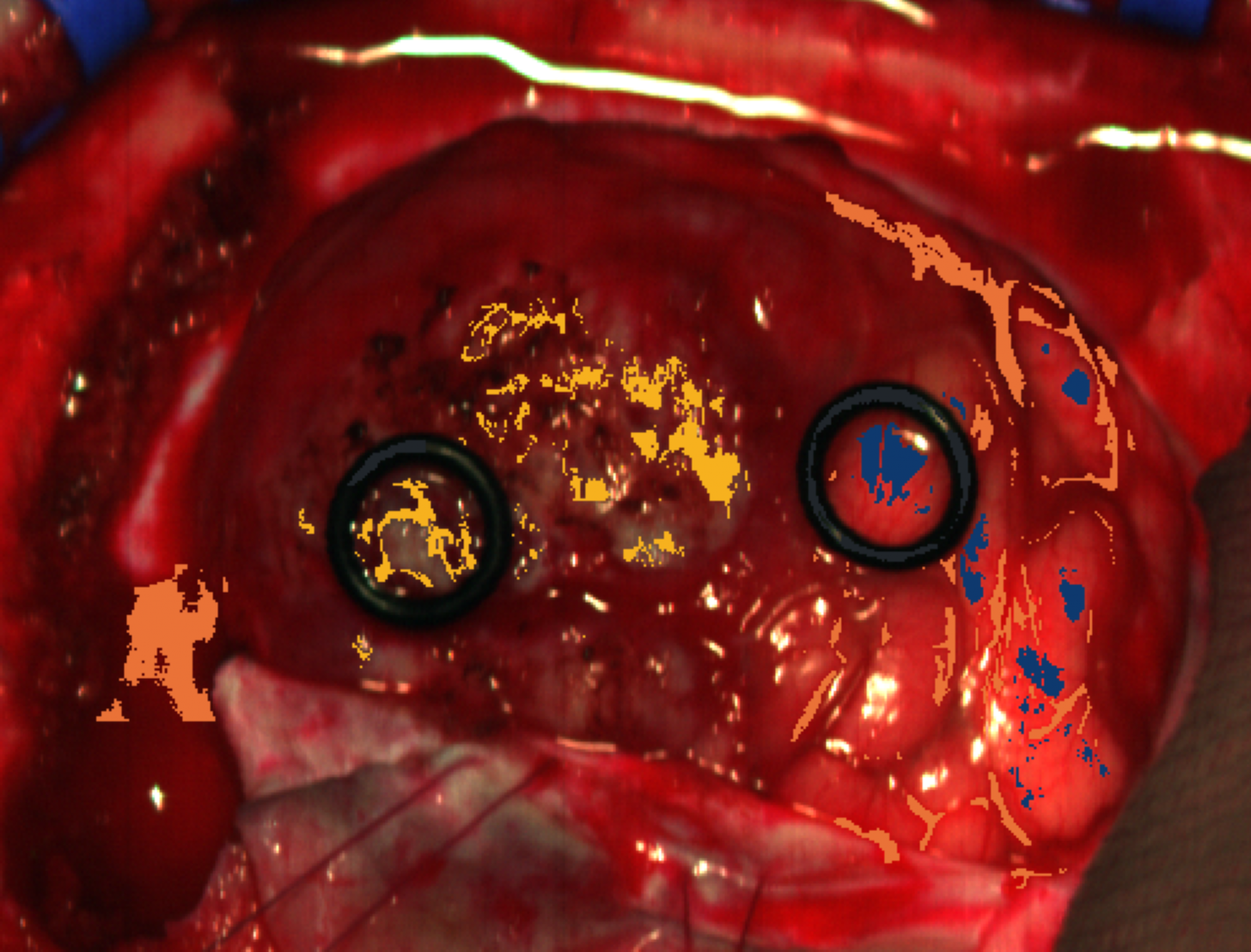} & \includegraphics[align=c, width=1.8cm]{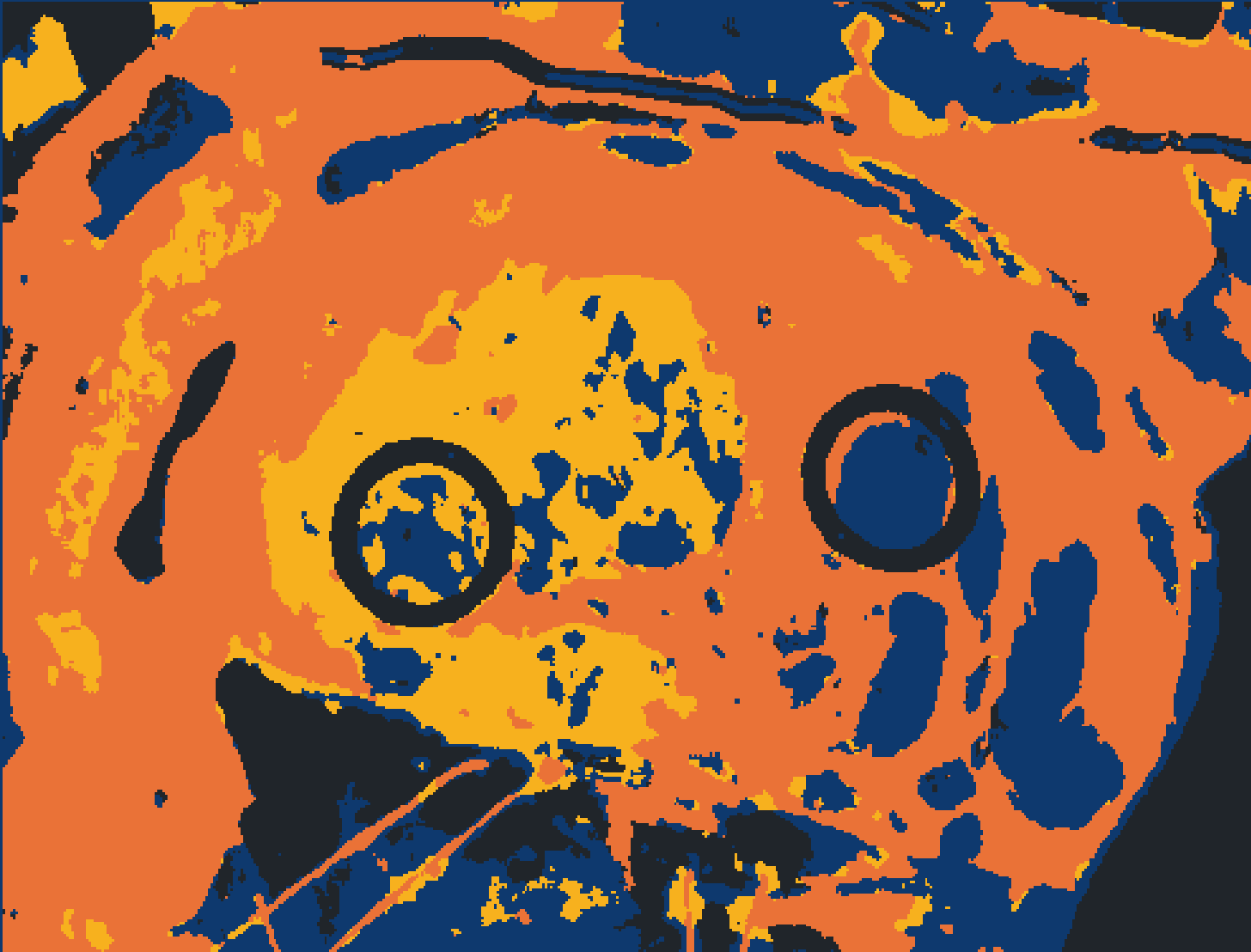} & \includegraphics[align=c, width=1.8cm]{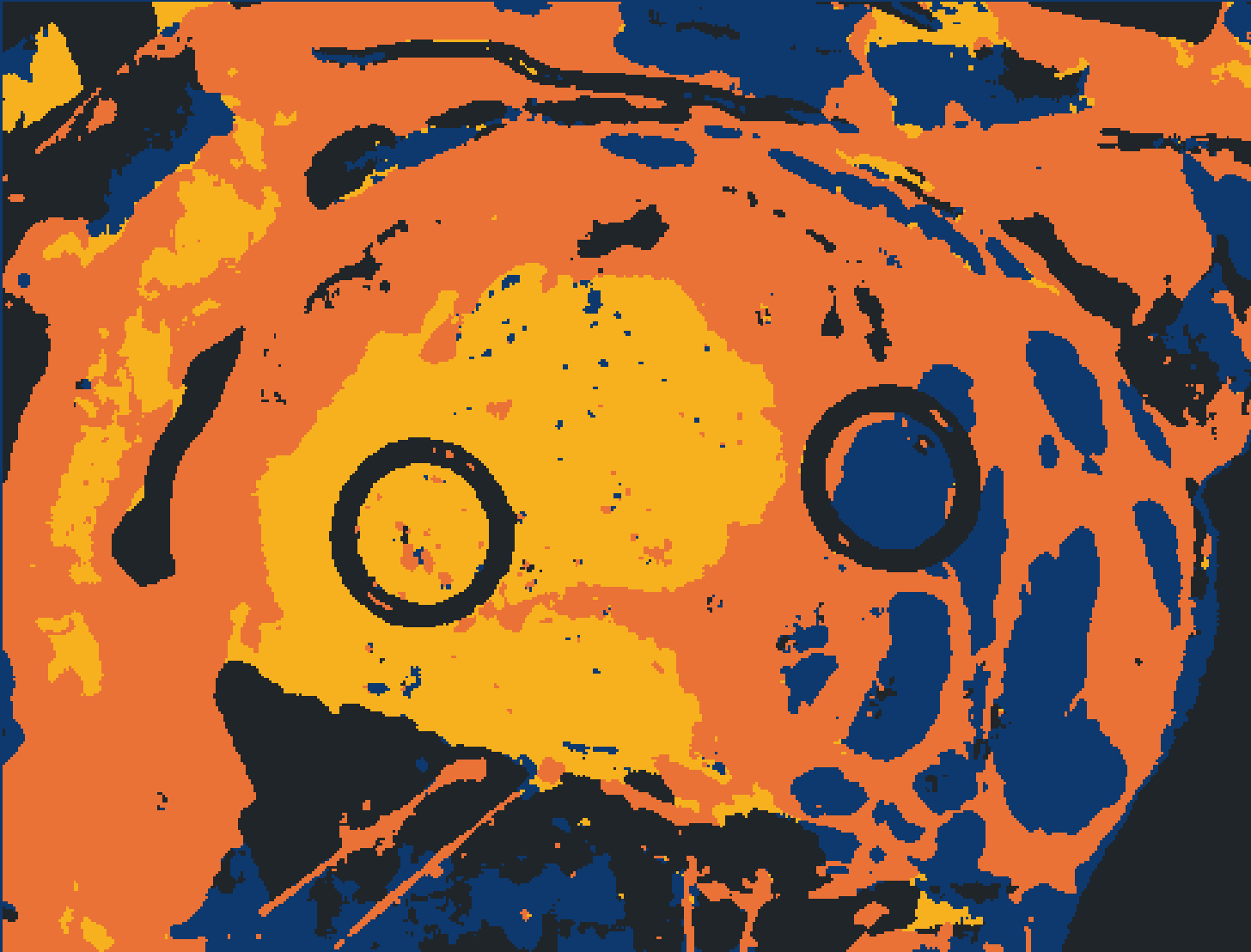} \\
            \end{tabular}
        \end{adjustbox}
    }
    \subfloat{
        \centering
        \begin{adjustbox}{valign=t}
            \setlength{\tabcolsep}{1pt}
            \begin{tabular}{@{} *{4}{c} @{}}
                & sRGB & BL Model & HM Model\\
                \verticalText{43-1} & \includegraphics[align=c, width=1.8cm]{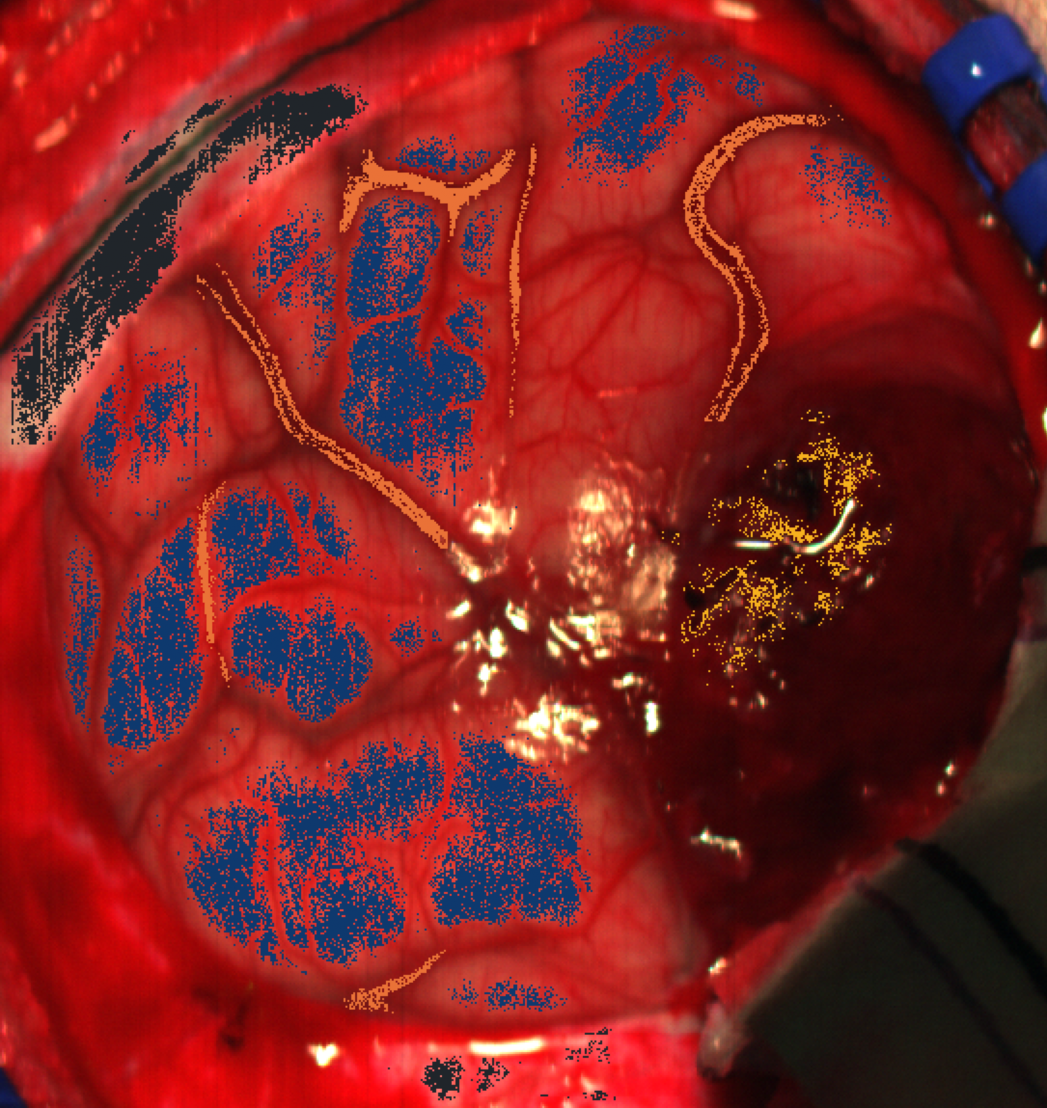} & \includegraphics[align=c, width=1.8cm]{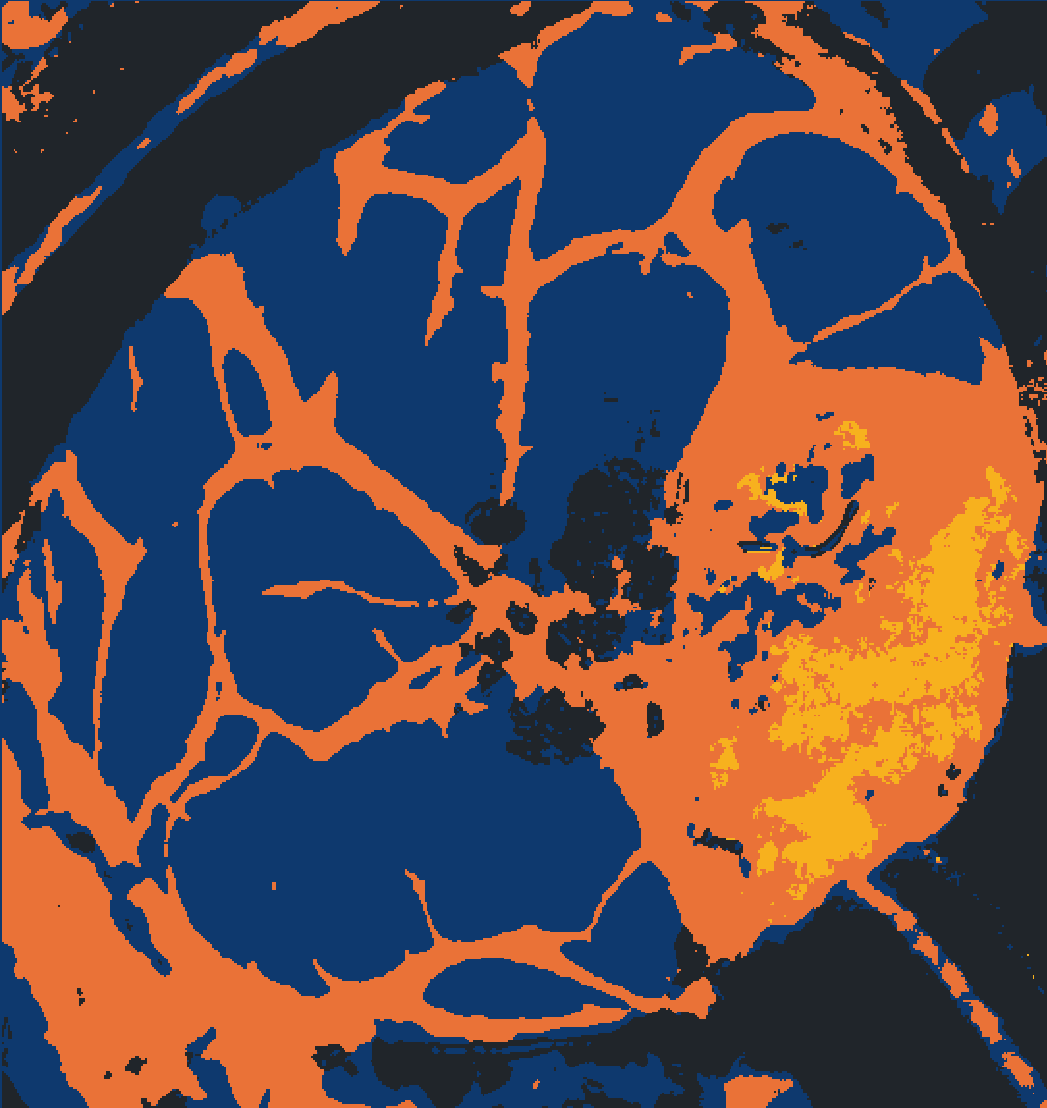} & \includegraphics[align=c, width=1.8cm]{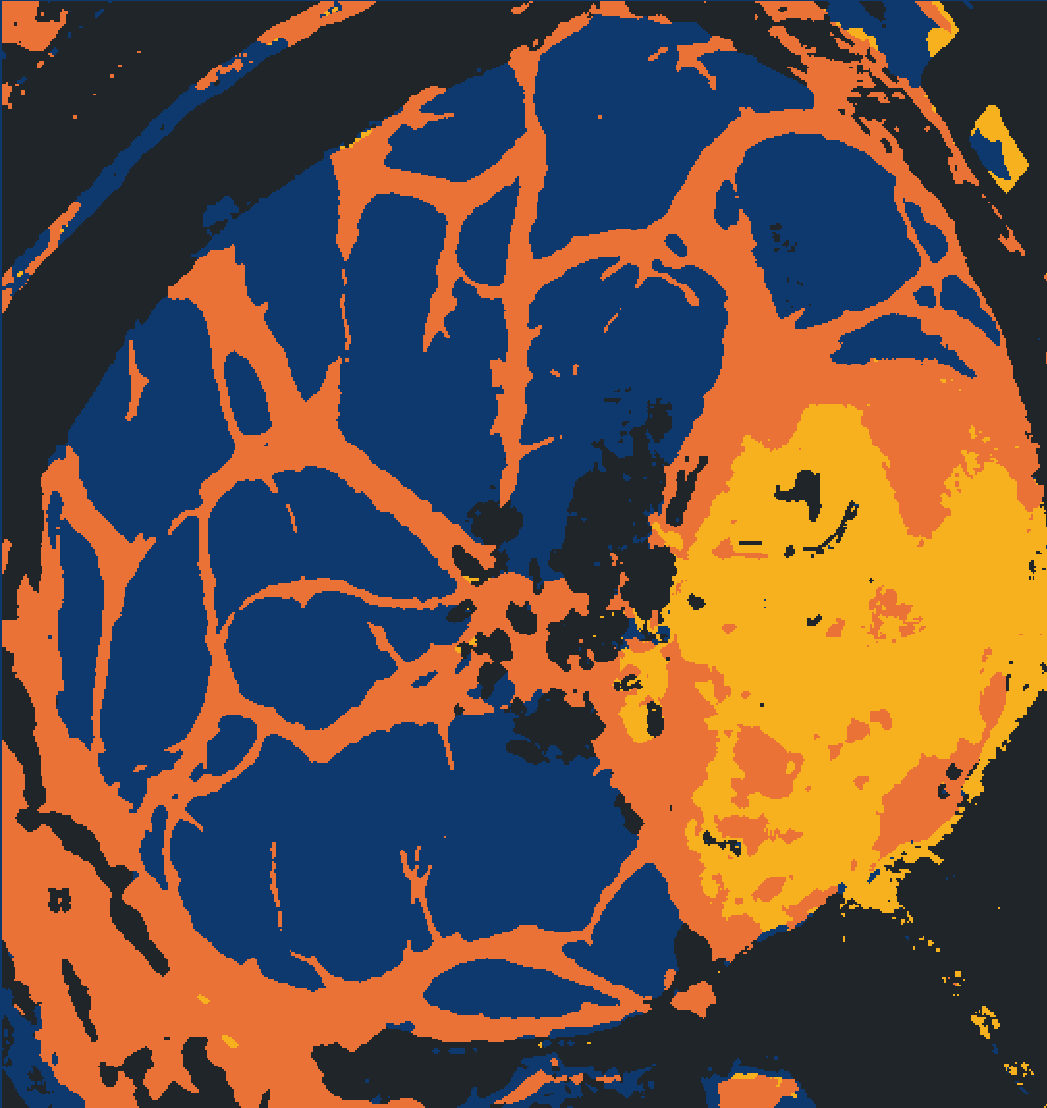} \\
            \end{tabular}
        \end{adjustbox}
    }
    \caption[]{Test image classification maps from the proposed HM (right) and baseline BL (middle) models, and the 
  sRGB image (left). \protect\linebreak \raisebox{0.6ex}{\colorbox{tum_blue_dark_2}{}}: normal tissue \raisebox{0.6ex}{\colorbox{tum_yellow}{}}: tumor tissue \raisebox{0.6ex}{\colorbox{tum_red}{}}: blood vessels \raisebox{0.6ex}{\colorbox{tum_grey_1}{}}: background.}
    \label{fig:classification_maps}
\end{figure}

Finally, Fig. \ref{fig:classification_maps} qualitatively demonstrates the observed trend of more cohesive segmentation obtained by the model with the proposed heatmaps. We attribute it to the fact that the molecular maps clearly outline physiologically different image parts further guiding the classifier to label identically pixels belonging to the same semantic class. 

A few important messages we would like to bring. The low F1 score for the tumor class is not caused by the nature of our method but the nature of the data: the objective difficulty of annotating heterogeneous and diffuse border glioma from HSI data and, as a result of it, the scarcity of such (biopsy-derived) annotations. A typical image in the Helicoid dataset (the only public glioma HSI dataset) contains up to 10$\%$ labeled pixels (see Fig. \ref{fig:classification_maps}, left subplots). This explains the low F1 score even for SOTA methods, because mispredictions have a disproportionately large impact on F1 when the number of labelled pixels is small.
However, qualitatively we observe a stark difference in segmentation obtained by the standard end-to-end segmentation method, Fig. 3 (middle), and the proposed approach (right). The latter provides consistently more cohesive segmentation covering tumor area. Quantitative difference in F1 is though less pronounced due to the mentioned limitations in annotation.

Next, in contrast to linear unmixing, the non-linear one is way more challenging to make work due to greater ill-posedness, no closed-form solution, and thus higher sensitivity to noise. For the Helicoid data of moderate SNR, this leads to unstable spectral unmixing, as we observed in our prior experiments. The same is found in the literature (see Fig. 6b in \cite{leon_hyperspectral_2023}, NEBEAE method), where unstable unmixing leads to radical degradation of performance on the segmentation task. That was the reason that brought us to develop the proposed methodology which is robust against the typical obstacles for non-linear unmixing.

\section{Conclusion}
In this study, we propose a novel spectral unmixing methodology for hyperspectral imaging of in-vivo brain tissue addressing limitations of standard unmixing. By using pseudo-endmembers derived from Monte Carlo simulations, our method enables a more accurate representation of light-matter interaction and enhances molecular biomarker extraction. Our results demonstrate that this approach improves the classification of brain tissue types reinforcing its potential applications in tissue characterization. While we employ a Monte Carlo technique in our work, the methodology is applicable to other physics simulators. We thus envision that the wider optics research community will adopt this methodology, approbating various complex light-matter modeling techniques.

Finally, we do not claim clinical readiness for tumor classification with the current F1 score accuracy by SOTA methods (the classification analysis was just one of several ways to quantitatively assess the utility of the extracted biomarkers by the proposed method). 
But we do believe that to achieve it, improvements in the methodology and improvements in the quality of ground truth annotation have to go hand in hand to mutually benefit from each other. Better model predictions will help to better annotate the images by experts in a semi-automated fashion, and vice versa, more and better annotated data will lead to superior models.

\section*{Acknowledgment}
The HyperProbe consortium has received funding from the European Union’s Horizon Europe Research (Grant No. 101071040). UCL is supported by UKRI (Grant No. 10048387). 

\bibliography{main1}
\bibliographystyle{spiebib}

\end{document}